\documentclass[%
 reprint,
 superscriptaddress,
 amsmath,amssymb,
 aps,
pra,
]{revtex4-2}
\usepackage[table]{xcolor}
\usepackage{graphicx}
\usepackage{dcolumn}
\usepackage{bm}

\usepackage[hidelinks]{hyperref}
\usepackage{orcidlink}
\usepackage{booktabs}
\usepackage[T1]{fontenc}

\begin{document}

\title{\textit{Ab initio} calculations of the electronic structure of Ac$^+$}

\newcommand\RUG{Faculty of Science and Engineering, Van Swinderen Institute for Particle Physics and Gravity, University of Groningen, 9747 AG Groningen, The Netherlands}

\author{Genevieve Geehan\orcidlink{0000-0003-3621-3588}}
\affiliation{\RUG}

\author{Marten Luit Reitsma\orcidlink{0000-0002-8255-7480}}
\affiliation{\RUG}

\author{Johan David Polet\orcidlink{0009-0000-2751-5635}}
\affiliation{\RUG}

\author{Mustapha Laatiaoui\orcidlink{0000-0003-0105-8303}}
\affiliation{GANIL, CEA/DRF-CNRS/IN2P3, B.P. 55027, 14076 Caen Cedex 05, France}

\author{Julian Berengut\orcidlink{0000-0002-7366-1091}}
\affiliation{School of Physics, University of New South Wales, Sydney NSW 2052, Australia}

\author{Anastasia Borschevsky\orcidlink{0000-0002-6558-1921}}
\affiliation{\RUG}

\date{\today}

\begin{abstract}
Accurate spectroscopic investigations of the heaviest elements are inherently challenging, due to their short lifetimes and low production yields. Success of such measurements requires both dedicated experimental techniques and strong theoretical support. Laser resonance chromatography (LRC) is a promising approach for heavy ion spectroscopy, in particularly for metals with low vapour pressure, such as actinium.  We have employed the state-of-the-art relativistic Fock space coupled cluster approach as well as the configuration interaction with many-body perturbation theory method to calculate the energy levels, the transition amplitudes, the branching ratios, and the hyperfine structure parameters of the lowest excited states in Ac$^+$. Knowledge of these properties is required for the design of experiments. Our calculations are in close agreement with experimental transition energies, leading us to expect a similar level of accuracy for the calculated hyperfine structure parameters. Based on these predictions, two possible experimental schemes are proposed for the planned LRC measurements.
\end{abstract}

\maketitle

\section{\label{sec:introduction}Introduction}
New experimental developments in accelerator physics and laser technology~\cite{CamMooPea16, BloLaaRae21, YanWanWil23} continue to push optical spectroscopy studies further into the uncharted territory of the heaviest elements. Spectroscopy studies help to elucidate the behaviour and electronic structure of these systems and probe the extreme effects of relativity. They can also provide access to nuclear properties, such as nuclear spins, moments, and charge radii, and can be used to test the predictive power of nuclear models.  The measurements are often driven by the need to obtain complementary information on the single-particle and collective properties of isotopes far from the valley of nuclear stability. Superheavy elements, however, remain a challenge for traditional spectroscopy methods, such as resonance ionization spectroscopy, due to the short half-lives and to the challenge of obtaining sufficient quantities~\cite{TrubertEtAl2003}. So far, the heaviest element investigated by means of optical spectroscopy is nobelium (No, Z=102). Sophisticated one-atom-at-a-time experiments were performed to measure the energy levels, the ionization potential, and the hyperfine structure of this unstable and short lived atom~\cite{LaatiaouiEtAl2016,ChhAckBac18,RaeAckBac18}. Great efforts have been made since then to extend this sophisticated type of spectroscopy to the next heavier element, lawrencium (Lr, Z=103)~\cite{MurRaeChh20, KahRaeEli21,AndBonJon25}, but so far no optical spectral lines have been recorded for this element.

Laser Resonance Chromatography (LRC) is proposed as an alternative method for heavy ion spectroscopy~\cite{LaatiaouiEtAl2020}. Due to its sensitivity it is promising for investigations of ions of heavy elements that can only be sourced in very small quantities~\cite{KimJanAry24}. The working principle of LRC leverages the drift time of ions through an inert gas, such as helium. The ion is scanned by a laser until a resonance is found, enabling electron shelving from the ground state to a metastable state. The interaction with the inert gas will be different for the different electronic configurations of the ion's ground and metastable state, which in turn will affect its transport properties and hence the drift time. In this way, accurate excitation energies of the ion can be determined, potentially including resolution of hyperfine structure components and thus providing access to nuclear properties.

For heavy and radioactive atoms, where the source material is available in small quantities, experiments must be as efficient as possible. Therefore, such measurements require high quality theoretical support to both guide the planning of the experiment and to avoid broad wavelength scans. 
On the one hand, if there are no prior experimental data available such as is the case for lawrencium, theoretical predictions allow spectroscopy to be carried out within a narrow search window, reducing the experiment run time~\cite{LaatiaouiEtAl2016,BloLaaRae21}.
On the other hand, theoretical support can also be crucial in the inevitable search for new avenues in spectroscopy by providing theoretical proof-of-principle and predictions of best-suited experimental boundary conditions, such as is the case for LRC on lutetium (Lu, Z=71)~\cite{LaatiaouiEtAl2020}, the lighter chemical homologue of lawrencium or actinium (Ac, Z=89).

While traditional spectroscopy of actinium is possible, it becomes challenging for some of its short lived isotopes that can be produced at Isotope Separation On-Line (ISOL) facilities~\cite{Jajčišinová2024}. For the spectroscopy of nobelium, which can be produced only in atom-at-a-time quantities, the radiation detection resonance ionization spectroscopy (RADRIS) technique has proven to be extremely sensitive. The method, however, faces some intrinsic limitations when it comes to spectroscopy of metals with low vapour pressures such as actinium~\cite{RaederEtAl2023}. In contrast, actinium should be an ideal candidate for LRC experiments. Ac$^+$, for instance, has a few low-lying excited states, which provide excellent excitation schemes that are suitable for LRC~\cite{NIST_ASD}. 

While the spectrum of Ac$^+$ is well documented~\cite{MegFreRom57}, extraction of the nuclear moments requires knowledge of the electronic hyperfine structure parameters of the states of interest. In addition, knowledge of the lifetimes of the different levels and the transitions strengths between them is important for setting up an effective optical pumping scheme. In order to be reliable and useful in experimental context, theoretical predictions should be based on high accuracy calculations. Furthermore, such predictions should be accompanied by quantifiable uncertainties. 

In this work we present high accuracy calculations of energies, lifetimes, and hyperfine structure parameters of six excited states of Ac$^{+}$. These states are the states of interest for the LRC scheme: $7s7p$ ($^3$P$^o_1$), $6d7p$ ($^1$P$^o_1$) and $6d7s$ ($^3$D$_1$), as well as three other levels, $6d7s$ ($^3$D$_2$) and ($^1$D$_2$) and $6d^2$ ($^3$F$_2$), for which existing experimental energies are compared with theory for validation of the method. Two relativistic high accuracy methods are used for the calculations in this work: the Fock space coupled cluster approach (FSCC)~\cite{visscher_2001_formulation, Pasteka2025} and configuration interaction method combined with many body perturbation theory (CI+MBPT)~\cite{dzuba96pra,Kahl&Berengut2019}. These methods have been extensively employed for calculations of  energies and properties of heavy elements and have demonstrated both high accuracy and considerable predictive power~\cite{KahBerLaa19,Pasteka2025}. In a number of recent theory-experiment publications, results obtained within both approaches are presented~\cite{RaeAckBac18, KahBerLaa19, BekBorHar19, GusRicRei20, KahRaeEli21}. In this work the CI+MBPT approach was augmented by including some chains of one-body core-valence correlations to all orders using Brueckner orbitals (CI+MBPT+Br); this augmentation improves the quality of the two-valence-electron wavefunction and hence the evaluation of matrix elements. Combination of the FSCC and the CI+MBPT+Br methods allows us to benefit from their complementary capabilities, provides an additional check on the accuracy and reliability of the results, and allows us to assign uncertainties based on the deviations between the methods. 

Recently, collinear spectroscopy measurements were performed on a number of neutron-rich actinium isotopes at the TRIUMF research facility~\cite{LiTeiLi2025}. The nuclear magnetic dipole and electric quadrupole moments of $^{225,226,228,229 }$Ac were extracted from the measured hyperfine structure of the $7s^2$ $^1$S$_0$ $\rightarrow$ $6d7p$ $^1$P$_1$ transition. However, a discrepancy in the values of the obtained nuclear magnetic moments with the previous measurements on neutral Ac~\cite{VerTeiRys19} motivates additional independent measurements for these isotopes.  

\section{\label{sec:comp_methods}Computational methods}

The FSCC calculations were carried out using the DIRAC24~\cite{DIRAC24} program, while CI+MBPT+Br calculations were performed within the AMBiT software~\cite{Kahl&Berengut2019}.
Calculations of excited-state lifetimes were executed only using the CI+MBPT+Br approach, as no efficient implementation for FSCC calculations of transition dipole moments is available in the DIRAC program.
Both methods are implemented within the framework of the  Dirac-Coulomb Hamiltonian~\cite{Suc80},
\begin{eqnarray}
H_{DC}= \displaystyle\sum\limits_{i}h_{D}(i)+\displaystyle\sum\limits_{i<j}(1/r_{ij}).
\label{eqHdcb}
\end{eqnarray}
Here, $h_D$ is the one electron Dirac Hamiltonian,
\begin{eqnarray}
h_{D}(i)=c\, \boldsymbol \alpha_{i}\cdot \mathbf{p}_{i}+c^{2}(\beta _{i}-1)+V_\textrm{nuc}(i),
\label{eqHd}
\end{eqnarray}
where $\bm{\alpha}$ and $\beta$ are the four-dimensional Dirac matrices. The nuclear potential $V_\textrm{nuc}$ accounts for the finite size of the nucleus. In the FSCC calculations, the Gaussian charge distribution was used~\cite{VisDya97}, while in the CI+MBPT+Br calculations we used the Fermi two-parameter charge distribution model. The choice of the finite nucleus model has been shown to have negligible effect on the calculated electronic properties~\cite{VisDya97}. The CI+MBPT+Br results also included the Breit and QED contributions. 
The following subsections provide an overview of the FSCC and the CI+MBPT+Br computational procedures.

\subsection{\label{sec:FSCC_Calculations}FSCC calculations}
The starting point for the multireference FSCC calculations is a closed-shell Ac$^{3+}$ $6p^6$  reference determinant, which was produced from a Dirac-Hartree-Fock (DHF) calculation. The FSCC calculations included single and double excitations, and the two-particle sector was used, where two electrons are added to the reference determinant to reach the Ac$^{+}$ states of interest. The electrons are allowed to occupy the orbitals specified by the model space $P$. By including the $7s$, $7p$, and $6d$ orbitals in $P$, all the states of interest can be reached. The model space was further increased to improve the correlation energy until convergence was reached.

In order to allow the use of the large model spaces while avoiding convergence issues of the coupled-cluster equations in the two-particle sector, the extrapolated intermediate Hamiltonian approach (XIH)~\cite{eliav_2005_extrapolated} was used. In this case the model space $P$ is divided into a main model space $P_m$ and an intermediate model space $P_i$, in which problematic energy denominators appearing the CC equations are shifted in energy. The excitation energies were obtained by extrapolating the results of calculations at different shift values to a shift of zero. The $7s$, $6d$ and $7p$ orbitals were included in the $P_m$ space, while the $P_i$ space covered the $5f$, $8p$, $6f$, $8s$ and $7d$ orbitals. All the electrons were correlated and virtual space cut-off was set at an energy of 100 a.u.

The uncontracted Dyall Gaussian basis sets ~\cite{dyall_2006_relativistic} were employed. The energies were calculated using the  valence basis sets augmented by adding three diffuse functions to each symmetry in an even-tempered fashion (t-aug-vXz); diffuse functions are important for capturing excitation energies correctly. We have corrected these results for the lack of inner-core correlating functions by taking the difference between a calculation that included these functions and the one that did not (see Table~\ref{tab:energies_correlation_correction} in the Appendix). This effect was obtained using v3z/ae3z basis sets, at t-aug-aev4z calculations were too computationally expensive. We have previously shown that the effect of the presence of core correlating functions is independent of the size and augmentation level of the basis set~\cite{Pasteka2025}.  

The HFS constants were obtained using the  all electron basis sets, which included correlating functions for all the valence and core shells (ae4z); for these properties, a high quality description of the nuclear region, provided by the core-correlating functions, is crucial. We have verified that the addition of diffuse functions to this basis set has a negligible effect on the calculated hyperfine structure parameters (see Table~\ref{tab:HFS_basis} in the Appendix for this computational investigation). The results were extrapolated to the complete basis set (CBS) limit employing the scheme of Martin~\cite{Martin:96} for correlation energies, 
\begin{equation}\label{eq:cbs_martin}
    E_N^\text{corr} = E_\text{CBS}^\text{corr} + \frac{A}{N^3}.
\end{equation}

Within the FSCC procedure, there is no need to extrapolate the DHF energy to the complete basis set limit. For transition energies, the CBS extrapolation of the DHF energy adds a constant shift to all level energies and thus cancels out for the transition energies. On the other hand, the closed shell ion state described by the DHF calculation does not contribute to the calculated hyperfine structure constants, due $J=0$ of this term. 

The  magnetic field at the nucleus, due to the electronic distribution, $A_0$, can be related to the measured magnetic dipole hyperfine coupling constant, $A$, by the following expression,

\begin{eqnarray}
    A = \frac{\mu}{I} A_0,
\end{eqnarray}
where $\mu$ is the nuclear magnetic moment and $I$ is the nuclear spin.

Similarly, the electric field gradient at the nucleus, $q_{zz}=\left<\frac{\partial^2 V}{\partial z^2}\right>$, can be related to the measured electric quadrupole hyperfine coupling constant $B$, 

\begin{eqnarray}
    B=eQq_{zz}, 
\end{eqnarray}
where $Q$ is the spectroscopic electric quadrupole moment of the nucleus. Combining the theoretical $A_0$ and $q_{zz}$ with the $A$ and $B$ parameters measured in a spectroscopy experiment allows for the extraction of the nuclear moments. 

For the calculation of the $A_0$ and $q_{zz}$ parameters, the finite field method was employed. The methodology discussed below is analogous to that of our earlier work for the HFS parameters of excited states in Sn~\cite{GusRicRei20} and Ge~\cite{Kanellakopoulos2020}; the relatively recent implementation described in Ref.~\cite{Haase2020} is used for the magnetic HFS parameters. In the finite field approach, the property of interest is added to the Hamiltonian as a perturbation,
\begin{equation}
 H = H_0 + \lambda H_{\mathrm{HFS}}^{A_0,q_{zz}}.
\end{equation}
Where $H_{\mathrm{HFS}}^{A_0,q_{zz}}$ is either the magnetic dipole HFS operator,
\begin{equation}\label{eq:Ahfs}
    H_{\mathrm{HFS}}^{A_0} = \sum_i \frac{(\boldsymbol{\alpha} \times \boldsymbol{r_i})}{r_i^3}, 
\end{equation}
or the electric field gradient operator,
\begin{equation}
        H_{HFS}^{q_{zz}} = \sum_i \frac{3z_i^2 - r_i^2}{r_i^5}.
\end{equation}
The expectation value of $H_{\mathrm{HFS}}^{A_0,q_{zz}}$ can be determined from the derivative of the energy with respect to $\lambda$ through the Hellman-Feynman theorem~\cite{Cohen1965,Pople1968,Monkhorst1977}, 
\begin{equation}
    \left< H_{\mathrm{HFS}}^{A_0,q_{zz}} \right> = \frac{E(\lambda) - E(-\lambda)}{2\lambda}.
\end{equation}
The finite field parameter $\lambda$ sets the strength of the perturbation and should be carefully chosen such that the energy change is large enough to overcome numerical noise and small enough to be in a linear regime, in this case $\lambda_{A_0} = 10^{-4}$ and $\lambda_{q_{zz}} = 5 \cdot 10^{-8}$.

\subsection{\label{subsec:CI+MBPT_Calculations} CI+MBPT+Br Method}
In the CI+MBPT method, valence-valence interactions are included in CI procedure, while core-valence interactions are treated using second-order MBPT~\cite{dzuba96pra}. In this work we additionally include some higher-order chains of one-body core-valence correlations using Brueckner orbitals, also implemented in the AMBiT code~\cite{Kahl&Berengut2019}. As with our FSCC calculations, the single-electron core orbitals are calculated in the $V^{N-2}$ (Ac$^{3+}$) Dirac-Fock potential.
The single electron Dirac-Fock operator $\hat h_D$ was modified to include the frequency-independent Breit interaction, Uehling potential, and self-energy (QED) corrections using the Flambaum-Ginges radiative potential approach~\cite{flambaum05pra,ginges16pra}.

An orbital basis set was formed by diagonalising a set of B-splines over the Dirac-Fock operator. This basis was used to calculate one-body MBPT diagrams at second order of perturbation theory, generating a matrix $\hat\Sigma_\kappa$ for each angular momentum. Diagonalising B-splines over the modified single-body operator
\begin{eqnarray}
    (\hat{h}_{D}+\hat{\Sigma}_\kappa)\varphi_{n\kappa}=\epsilon_{n\kappa}\varphi_{n\kappa}
    \label{eqn:modOperator}
\end{eqnarray}
gives an orthonormal set of single-electron Brueckner orbitals $\varphi_{n\kappa}$ that include the effects of one-body core-valence correlations~\cite{dzuba85jpb,dzuba87jpb}. These valence and virtual orbitals are then used throughout the CI+MBPT calculation.

Interactions between the two valence electrons are included using CI. The valence basis set includes orbitals with $l\leq 4$ and $n\leq20$, which is converged in transition energy to within $\sim 1\%$. 
The two-body core-valence interactions are treated at second order according to the usual CI+MBPT methodology by modifying the two-electron Slater integrals that are used in CI~\cite{dzuba96pra}. The diagrams include excitations into virtual orbitals with $l\leq 8$ and $n\leq30$. Convergence of the MBPT basis is also below $1\%$ for transition energies.

The CI+MBPT+Br method differs from the standard CI+MBPT only by the treatment of one-body core-valence correlations. Yet the resulting wavefunctions better incorporate core-correlation effects and are therefore more suited to the calculation of matrix elements. The matrix elements required for transition rates and hyperfine structure parameters were calculated by taking the expectation value of these wavefunctions over a modified operator that additionally includes corrections of the random-phase approximation (RPA)~\cite{dzuba87jpb}. RPA models the polarisation of the core due to the external operator (e.g. nuclear magnetic moment or applied electric field) that are not otherwise captured.

We also note that in Eq.~\ref{eq:Ahfs}, the effect from a finite nuclear magnetization distribution (the Bohr-Weisskopf effect) is neglected. It has been shown previously that this effect can alter the magnetic hyperfine parameter on the order of ten percent for heavy elements~\cite{ginges2017ground, roberts2020nuclear}. We have performed our CI+MBPT+Br calculations both with and without this effect, and found that it changes the values of $A_0$ by around 3\% percent and $q_{zz}$ by around 1\%: less than the difference with our FSCC calculations (which do not include this effect).

\section{Results}
\subsection{\label{sec:energies}Energies}
The energies of the excited states of interest, calculated by both FSCC and the CI+MBPT+Br approaches are presented in Table~\ref{tab:Energies} and compared to the available experimental values from the NIST database~\cite{NIST2005}.
\begin{table}[b]
    \caption{Excited state energies (cm$^{-1}$) of Ac$^+$ obtained with the FSCC and CI+MBPT+Br method and compared to experiment.
    \label{tab:Energies}}
    \begin{ruledtabular}
    \begin{tabular}{@{}llccc@{}}
    && \multicolumn{3}{c}{\textrm{Energy}}\\
    \cline{3-5}
    \multicolumn{2}{c}{\textrm{State}}&  \textrm{FSCC}      & \textrm{CI+MBPT+Br} &\textrm{Expt.\cite{NIST2005}}\\
    \colrule
    $6d7s$  & $^{3}$D$_{1}$     & 5190  & 4942  & 4740  \\
            & $^{3}$D$_{2}$     & 5726  & 5544  & 5267  \\
            & $^{1}$D$_{2}$     & 9630  & 9709  & 9088  \\
    $6d^2$  & $^{3}$F$_{2}$     & 14006 & 13828 & 13236 \\
    $7s7p$  & $^{3}$P$^{o}_{1}$ & 22153 & 23665 & 22181 \\
    $6d7p$  & $^{1}$P$^{o}_{1}$ & 29546 & 30294 & 29250 \\
    \end{tabular}
    \end{ruledtabular}
\end{table}
The energies predicted by both approaches show good agreement with the experimental values, with average error of around $\sim5\%$ for both methods, lending confidence to the predictions of the lifetimes and the hyperfine structure constants.

\subsection{\label{sec:lifetimes}Lifetimes}
Lifetimes are given by the inverse of the sum of rates of spontaneous emission. These partial rates are derived from the relation
\begin{eqnarray} \label{equ:rate}
    \Gamma_K=\frac{\kappa(2K + 2)(2K+1)}{K[(2K+1)!!]^2(2J_i+1)} \left(\frac{\omega}{c}\right)^{2K+1} S_K,
\end{eqnarray}
where $S_K$ is the transition line strength (the square of the transition matrix element), $K$ is the multipolarity of transition, $\kappa=1$ for electric and $\kappa=1/4c^2$ for magnetic multipole transitions, $J_i$ is the total angular momentum of the initial state and $\omega$ is the frequency of the transition. Experimental energies and matrix elements from CI+MBPT+Br calculations were used in the calculations of the decay rates and, hence, lifetimes. The calculated partial transition rates are given in Table~\ref{tab:Rates} and compared to the predictions of Kramida~\cite{Kramida2022} and Roberts et al.~\cite{roberts2014strongly}.
\begin{table*}[h!t]
    \caption{\label{tab:Rates}%
    Calculated transition rates of the states of interest, with comparison to the predictions of Ref.~\cite{Kramida2022} and Ref.~\cite{roberts2014strongly}.}
    \begin{ruledtabular}
    \begin{tabular}{@{}l|cclllll@{}}
    &{\textrm{Type}}&$\lambda$ (nm)&\multicolumn{3}{c}{\textrm{$\Gamma$ (s$^{-1}$)}}\\
    \textrm{Transition}&  &Ref.~\cite{NIST2005}&This work&Referenced\\
    \colrule
    \noalign{\vspace{2pt}}
    $6d7s$ $^3$D$_{1}$ $\rightarrow$ $7s^2$ $^1$S$_0$ & M1 &{2.11$\times10^{3}$}&{4.93$\times10^{-6}$}&1.1$\times10^{-14}$ \cite{Kramida2022}\\
    
    $6d7s$ $^3$D$_{2}$ $\rightarrow$ $7s^2$ $^1$S$_0$ & E2 & {1.90$\times10^{3}$}&{2.37$\times10^{-3}$}&3.2$\times10^{-3}$ \cite{Kramida2022}\\
    
    $6d7s$ $^3$D$_{2}$ $\rightarrow$ $6d7s$ $^3$D$_1$ & M1 & {1.89$\times10^{4}$}&{3.11$\times10^{-3}$}&3.11$\times10^{-3}$ \cite{Kramida2022}\\
    
    $6d7s$ $^1$D$_{2}$ $\rightarrow$ $7s^2$ $^1$S$_0$ & E2 &  {1.10$\times10^{3}$}&{2.53$\times10^{-1}$}&3.2$\times10^{-1}$ \cite{Kramida2022}\\
    
    $6d7s$ $^1$D$_{2}$ $\rightarrow$ $6d7s$ $^3$D$_1$ & {M1}  &{2.30$\times10^{3}$}&{2.62$\times10^{-1}$}&2.55$\times10^{-1}$ \cite{Kramida2022}\\
    
    $6d7s$ $^1$D$_{2}$ $\rightarrow$ $6d7s$ $^3$D$_2$ &{M1(97\%)+E2(3\%)}&{2.62$\times10^{3}$}&{2.97$\times10^{-2}$}&2.91$\times10^{-2}$ \cite{Kramida2022}\\
    
    $6d^2$ $^3$F$_{2}$ $\rightarrow$ $7s^2$ $^1$S$_0$ &{E2}&{7.56$\times10^{2}$}&{1.33$\times10^{-1}$}&1.0$\times10^{-1}$ \cite{Kramida2022}\\
    
    $6d^2$ $^3$F$_{2}$ $\rightarrow$ $6d7s$ $^3$D$_1$ &{M1(25\%)+E2(75\%)}&{1.18$\times10^{3}$}&{2.60$\times10^{-1}$}&2.5$\times10^{-1}$ \cite{Kramida2022}\\
    
    $6d^2$ $^3$F$_{2}$ $\rightarrow$ $6d7s$ $^3$D$_2$ &{M1(3\%)+E2(97\%)} &{1.25$\times10^{3}$}&{8.83$\times10^{-2}$}&1.2$\times10^{-1}$ \cite{Kramida2022}\\
    
    $6d^2$ $^3$F$_{2}$ $\rightarrow$ $6d7s$ $^1$D$_2$ &{M1}&{2.41$\times10^{3}$}&{5.30$\times10^{-2}$}&2.9$\times10^{-2}$ \cite{Kramida2022}\\
    
    $7s7p$ $^3$P$^{o}_{1}$ $\rightarrow$ $7s^2$ $^1$S$_0$ &{E1}&{4.51$\times10^{2}$}&{2.17$\times10^7$}&2.13$\times10^7$ \cite{roberts2014strongly}\\
    
    $7s7p$ $^3$P$^{o}_{1}$ $\rightarrow$ $6d7s$ $^3$D$_1$ &{E1}&{5.73$\times10^{2}$}&{9.66$\times10^6$}&1.04$\times10^7$ \cite{roberts2014strongly}\\ 
    
    $7s7p$ $^3$P$^{o}_{1}$ $\rightarrow$ $6d7s$ $^3$D$_2$ &{E1}&{5.91$\times10^{2}$}&{1.60$\times10^7$}&1.58$\times10^7$ \cite{roberts2014strongly}\\
    
    $6d7p$ $^1$P$^{o}_{1}$ $\rightarrow$ $7s^2$ $^1$S$_0$ &{E1}&{3.42$\times10^{2}$}&{9.46$\times10^{7}$}&1.11$\times10^{8}$ \cite{Kramida2022}\\
    
    $6d7p$ $^1$P$^{o}_{1}$ $\rightarrow$ $6d7s$ $^3$D$_1$ &{E1}&{4.08$\times10^{2}$}&{7.13$\times10^{6}$}&1.09$\times10^{7}$ \cite{Kramida2022}\\
    
    $6d7p$ $^1$P$^{o}_{1}$ $\rightarrow$ $6d7s$ $^3$D$_2$ &{E1}&{4.17$\times10^{2}$}&{6.33$\times10^{7}$}&7.0$\times10^{7}$ \cite{Kramida2022}\\
    
    $6d7p$ $^1$P$^{o}_{1}$ $\rightarrow$ $6d7s$ $^1$D$_2$ &{E1}&{4.96$\times10^{2}$}&{1.88$\times10^{7}$}&1.20$\times10^{7}$ \cite{Kramida2022}\\
    
    $6d7p$ $^1$P$^{o}_{1}$ $\rightarrow$ $6d^2$ $^3$F$_2$ &{E1}&{6.24$\times10^{2}$}&{1.23$\times10^{7}$}&1.86$\times10^{7}$ \cite{Kramida2022}\\
    \end{tabular}
    \end{ruledtabular}
\end{table*}
We have followed the convention used by~\cite{Kramida2022} to label the different transition types, with mixed types having more than 2\% contribution of a second multipolarity.
There is good agreement for all the calculated transition rates except for the $6d7s$ ($^3$D$_1$) $\rightarrow$ $7s^2$ ($^1$S$_0$) transition, where the CI+MBPT+Br result is higher than the predicted rate from Ref.~\cite{Kramida2022} by several orders of magnitude. However, this discrepancy is not significant in the context of the LRC experiment, as the measurement does not rely on the precise knowledge of the lifetime of the $6d7s$ ($^3$D$_1$) state. This state should just be long lived enough to be detected, which is the case for both lifetime predictions.

The calculated spontaneous decay transition rates and lifetimes of the states of interest are shown in Table~\ref{tab:Rates_Lifetimes}. Spontaneous decay rates are taken as the sum of decay rates of all available decay pathways from the state of interest and lifetimes are the inverse of the spontaneous decay rates. We estimate the accuracy of these rates at within $\sim\! 10\%$ based on the difference between the CI+MBPT+Br and CI+MBPT methods (i.e. modification of the single body operator shown in equation~\ref{eqn:modOperator}).

\begin{table}[b]
    \caption{Radiative lifetimes and total spontaneous decay rates, $\sum\Gamma$, for the exited states of interest. We estimate the accuracy at $\sim 10\%$; see text for details.}\label{tab:Rates_Lifetimes}
    \begin{ruledtabular}
    \begin{tabular}{@{}ccllc@{}}
    \multicolumn{2}{c}{\textrm{Level}} & \textrm{Lifetime (s)}& $\sum\Gamma$ (s$^{-1}$) &\textrm{Dominant Channel}\\
    \colrule
    $6d7s$ &$^{3}$D$_{1}$ & {2.0$\times10^{5}$} &{4.9$\times10^{-6}$}& \textrm{M1}\\
        &$^{3}$D$_{2}$ & {1.8$\times10^{2}$} &{5.5$\times10^{-3}$}& \textrm{M1(57\%) \& E2(43\%)}\\
        &$^{1}$D$_{2}$ & {1.8} &{5.6$\times10^{-1}$}& \textrm{M1(55\%) \& E2(45\%)}\\
        $6d^2$ &$^{3}$F$_{2}$ & {1.8} &{5.5$\times10^{-1}$}& \textrm{M1(25\%) \& E2(75\%)}\\
        $7s7p$ &$^{3}$P$^{o}_{1}$ & {2.1$\times10^{-8}$} &{4.75$\times10^{7}$}& \textrm{E1}\\
        $6d7p$ &$^{1}$P$^{o}_{1}$ & {5.1$\times10^{-9}$} &{2.0$\times10^{8}$}& \textrm{E1}\\
    \end{tabular}
    \end{ruledtabular}
\end{table}

The two odd states $7s7p\ ^{3}$P$^{o}_{1}$ and $6d7p\ ^{1}$P$^{o}_{1}$ were found to be short lived with lifetimes of $\sim10^{-9}-10^{-8}$~s, while the $6d7s\ ^{3}$D$_{1}$ level was found to be metastable with a lifetime of $10^5$~s. 

\subsection{\label{sec:HFS}Hyperfine Structure Parameters}

The calculated values of $A_0$ and $q_{zz}$, obtained using the FSCC and the CI+MBPT+Br methods  are displayed in Table~\ref{tab:HFS}.
\begin{table}[h!t]
    \caption{HFS constants ($A_o$ and $q_{zz}$) of Ac$^+$ excited states calculated with the FSCC and CI+MBPT+Br methods. \label{tab:HFS}}
    \begin{ruledtabular}
    \begin{tabular}{@{}cccccccc@{}}
    \multicolumn{2}{c}{\textrm{Level}} &&\multicolumn{2}{c}{$A_0$ (MHz)}&&\multicolumn{2}{c}{$q_{zz}$ (MHz/b)}\\
                 & &&\textrm{FSCC}&\textrm{CI+MBPT+Br}                &&\textrm{FSCC}&\textrm{CI+MBPT+Br}\\
    \hline
    $6d7s$ & $^{3}$D$_{1}$     && -5265 &{-5233}
                                     && 231   &{196}\\
                 & $^{3}$D$_{2}$     && 4974  &{4728}
                                     && 369   &{345}\\
                 & $^{1}$D$_{2}$     && -2422 &{-2506}
                                     && 391   &{449}\\
    $6d^2$       & $^{3}$F$_{2}$     && 753   &{850}
                                     && 176   &{148}\\
    $7s7p$ & $^{3}$P$^{o}_{1}$ && 12471 &{12790}
                                     && -542  &{-424}\\
    $6d7p$ & $^{1}$P$^{o}_{1}$ && -2331 &{-2189}
                                     && 283   &{206}\\
    \end{tabular}
    \end{ruledtabular}
\end{table}

The calculated $A_0$ parameters are in good agreement between the two methods, with discrepancies of well below 5$\%$ in most cases, and the largest difference of 13$\%$ for the $^3$F$_2$ level. In contrast, larger discrepancies can be seen for $q_{zz}$. For the even states, the discrepancies range from $6 - 15 \%$ and for the odd states they are even greater. The physical origin of the discrepancy between the two methods is not easy to disentangle given that the approaches are very different. The effect of core-polarization in the CI+MBPT+Br method (treated in the random-phase approximation) is up to 60\%, so the discrepancy may well be due to the more complete treatment of core polarization in the FSCC method.

The $q_{zz}$ and $A_0$ parameters for the $^1$P$^o_1$ state were also calculated using multiconfigurational Dirac Hartree Fock (MCDHF) approach in Ref.~\cite{LiTeiLi2025} and used there to extract the nuclear moments from measurements. The values reported in Ref.~\cite{LiTeiLi2025} are $-1890(190)$\,MHz for $A_0$ and 173(17)\,MHz for $q_{zz}$, which are somewhat lower than the current predictions.
Given the hyperfine parameters listed in Table~\ref{tab:HFS}, a relative mobility difference of more than 7$\%$ between the ground state and the metastable $^3$D states, and an LRC setup with chromatography performance comparable to that reported in Ref.~\cite{KimJanAry24}, one can expect to resolve all three hyperfine peaks of both the $^3$P$^{o}_1$ and $^1$P$^{o}_1$ states.

\section{Potential optical pumping schemes for LRC}\label{sec:Schemes}

The combination of the short lifetimes of the $7s7p\ ^{3}$P$^{o}_{1}$ and $6d7p\ ^{1}$P$^{o}_{1}$ states, along with the long-lived metastable $6d7s\ ^{3}$D$_{1}$ state provides an ideal excitation scheme to probe either of the odd states with LRC. 

\begin{figure}[hbtp]
    \includegraphics[width=75mm, trim=10mm 0 0 0]{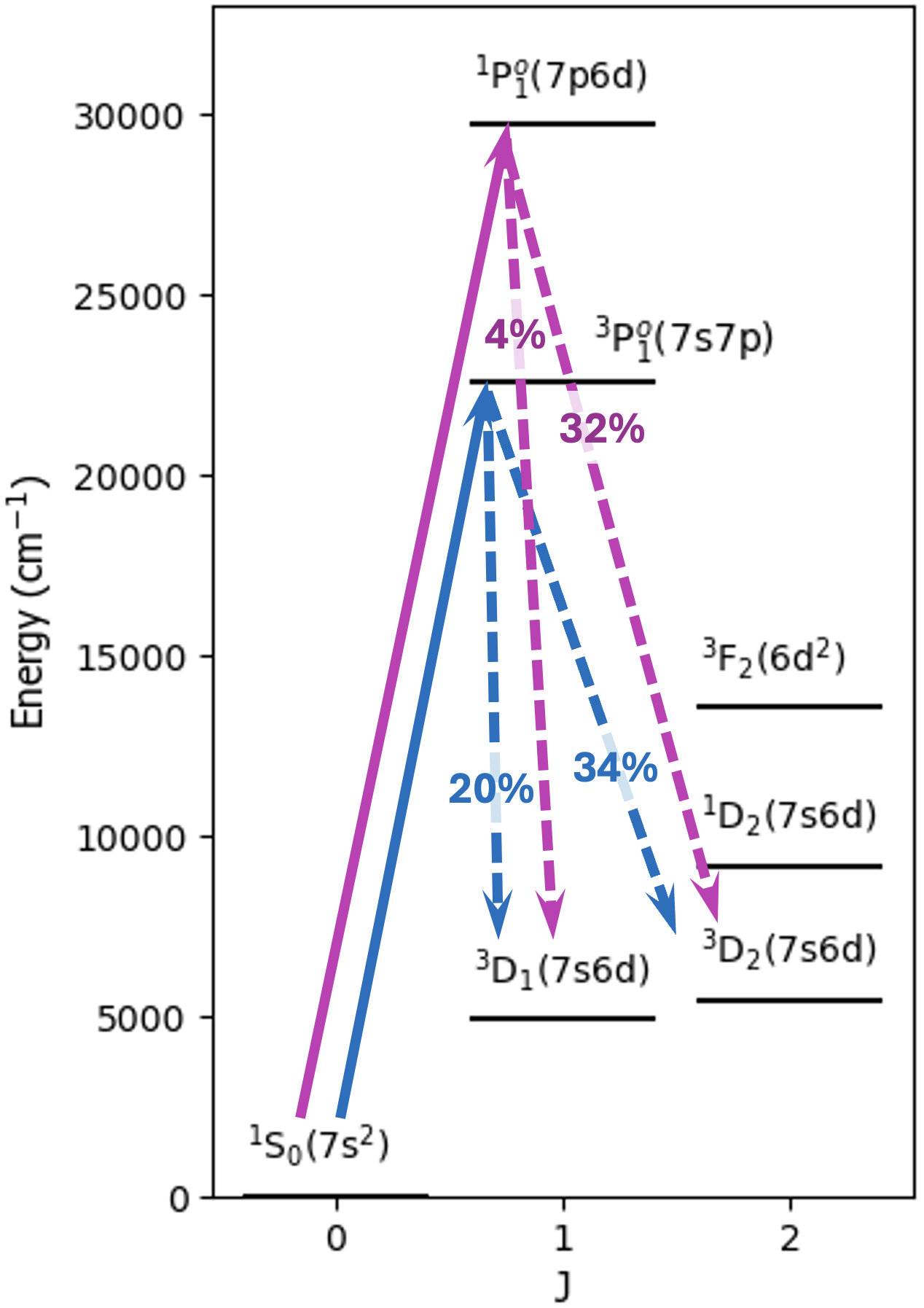}
    \caption{\label{fig:Ac+_levels} Level scheme for the lowest levels in Ac$^+$. Two possible optical pumping schemes are indicated by solid and dashed arrows of different colors. The percentage numbers given are the calculated branching fractions to metastable states $^3$D$_1$ and $^3$D$_2$.}
\end{figure}

Figure~\ref{fig:Ac+_levels} shows the structure of the lowest-lying levels in Ac$^+$, together with two potential optical pumping schemes so far identified for future laser resonance chromatography experiments.
The first one involves ground-state laser excitations to the $7s7p$ $^3$P$^{o}_1$ state in the visible spectral range at about $451\,$nm whereas the second involves resonant optical pumping via the bright $6d7p$ $^1$P$^{o}_1$ state in the UV range at about $342\,$nm. The scanning laser will scan close to the resonant frequency of either of these states. Due to the high transition rates, either of the odd levels considered here will be efficiently excited when the laser hits the resonant frequency. The ions will immediately decay from the short-lived excited states into the metastable $6d7s\ ^{3}$D states. It is these states that will be detected as a peak in the drift time spectra. 

Both ground-state transitions are very strong and the radiative feeding of the $6d7s\,^3$D$_2$ and $6d7s\,^3$D$_1$ metastable states is very efficient. According to first estimates based on a rate equation model for optical pumping, as developed in Ref.~\cite{Visentin:2024}, we obtain a combined occupation probability for the two metastable states of more than $98$\% after ten laser pulse exposures of $10\,$ns duration. As in Ref.~\cite{Visentin:2024}, we assumed laser exposures every $100\,$µs, an energy density of the laser radiation of $2.5\,$µJ/cm$^2$, and a four-level system, consisting of the ground state, one of the bright states to be probed, and the two 6d7s$\,^3$D$_2$ and 6d7s$\,^3$D$_1$ metastable states. These high occupation probabilities seem to vary only slightly (within $1$\%) with the choice of the Einstein $A$ coefficients for the different transitions presented in Table~\ref{tab:Rates}. Therefore, both proposed pumping schemes should readily enable laser resonance chromatography on Ac$^+$.

\section{\label{sec:conclusion}Conclusion}
High accuracy theoretical predictions have been made of the energies, lifetimes, and hyperfine structure parameters of the excited states of Ac$^+$, focusing on the states of interest for LRC spectroscopy. Two state-of-the-art relativistic methods were used, namely the FSCC and the CI+MBPT+Br approaches. The calculated energies were in good agreement between the two methods and with the experimental values, lending confidence to our predictions for the HFS constants, where no experiment is available. Based on the predicted atomic properties, two potential experimental schemes are proposed which will allow resolution of the hyperfine peaks of the $7s7p$ $^3$P$^{o}_1$ and $6d7p$ $^1$P$^{o}_1$ levels of interest.

\begin{acknowledgments}
This project has received funding from the European Union’s Horizon
2020 research and innovation programme under grant agreement No 861198–LISA–H2020-MSCA-ITN-2019. AB and MLR thank the Dutch Research Council (NWO) project number Vi.Vidi.192.088. We thank the Center for Information Technology of the University of Groningen for their support and for providing access to the Hábrók high performance computing cluster. M.L. gratefully acknowledges the support from GANIL, CNRS-IN2P3, and CEA-DRF. 
\end{acknowledgments}

\appendix*
\section{Computational investigations}\label{appendix:compinvestigation}

The effect of inner-core correlating functions in the basis set are presented in Table~\ref{tab:energies_correlation_correction}.
\begin{table}[b]
    \caption{Excited state energies (cm$^{-1}$) of Ac$^+$ obtained with the FSCC approach and two different computational settings: 1) v3z basis set, correlating all the electrons and with a virtual cutoff of 100~a.u.; 2) ae3z basis set, correlating all the electrons and a virtual space cutoff of 500~a.u.\label{tab:energies_correlation_correction}}
    \begin{ruledtabular}
    \begin{tabular}{lcccc}
  \multicolumn{2}{c}{\textrm{Level}}  &\textrm{Setting 1}      & \textrm{Setting 2} &\textrm{Diff.}\\
    \colrule
    $6d7s$ & $^{3}$D$_{1}$      &  6233 & 6311 &88    \\
                  & $^{3}$D$_{2}$     & 6765 & 6847  &81   \\
                  & $^{1}$D$_{2}$     & 10773 & 10851&78    \\
    $6d^2$        & $^{3}$F$_{2}$     & 16008  & 16188  &180 \\
    $7s7p$  & $^{3}$P$^{o}_{1}$ &  22219 & 22278 & 59\\
    $6d7p$  & $^{1}$P$^{o}_{1}$ &   30305& 30437 &131  \\
    \end{tabular}
    \end{ruledtabular}
\end{table}

Table~\ref{tab:HFS_basis} demonstrates that the HFS constants are not sensitive to the inclusion of additional diffuse functions to the basis set.
\begin{table}[ht]
    \caption{HFS constants ($A_o$ and $q_{zz}$) of Ac$^+$ excited states calculated using the dyall.v4z and the d-aug-dyall.v4z basis sets. The calculations were performed using the same computational settings as described in the body of the paper. \label{tab:HFS_basis}}
    \begin{ruledtabular}
    \begin{tabular}{@{}cccccccc@{}}
    \multicolumn{2}{c}{\textrm{Level}} &&\multicolumn{2}{c}{$A_0$ (MHz)}&&\multicolumn{2}{c}{$q_{zz}$ (MHz/b)}\\
                 & &&\textrm{v4z}&\textrm{d-aug-v4z}                &&\textrm{v4z}&\textrm{d-aug-v4z}\\
    \hline
    $6d7s$ & $^{3}$D$_{1}$     && -5045 &{-5045}
                                     && 230   &{230}\\
                 & $^{3}$D$_{2}$     && 4807  &{4811}
                                     && 378 &{378}\\
                 & $^{1}$D$_{2}$     && -2384&{-23887}
                                     && 409  &{410}\\
    $6d^2$       & $^{3}$F$_{2}$     && 774   &{774}
                                     && 178   &{178}\\
    $7s7p$ & $^{3}$P$^{o}_{1}$ && 11023 &{11022}
                                     && -619  &{-619}\\
    $6d7p$ & $^{1}$P$^{o}_{1}$ && -2707 &{-2720}
                                     && 438   &{440}\\
    \end{tabular}
    \end{ruledtabular}
\end{table}

\bibliography{references.bib}

\end{document}